\documentclass[11pt,twocolumn]{article}
\usepackage[utf8]{inputenc}
\usepackage{amsmath,amssymb,amsthm}
\usepackage{graphicx}
\usepackage{hyperref}
\usepackage{algorithm}
\usepackage{algpseudocode}
\usepackage{booktabs}
\usepackage{natbib}

\title{Privacy-Preserving Cohort Analytics for Personalized Health Platforms: A Differentially Private Framework with Stochastic Risk Modeling}

\author{Richik Chakraborty, Lawrence Liu, and Syed Hasnain}

\date{}

\begin{document}

\maketitle

\begin{abstract}
Personalized health analytics increasingly rely on population benchmarks to provide contextual insights such as ''How do I compare to others like me?'' However, cohort-based aggregation of health data introduces nontrivial privacy risks, particularly in interactive and longitudinal digital platforms. Existing privacy frameworks such as $k$-anonymity and differential privacy provide essential but largely static guarantees that do not fully capture the cumulative, distributional, and tail-dominated nature of re-identification risk in deployed systems.

In this work, we present a privacy-preserving cohort analytics framework that combines deterministic cohort constraints, differential privacy mechanisms, and synthetic baseline generation to enable personalized population comparisons while maintaining strong privacy protections. We further introduce a stochastic risk modeling approach that treats re-identification risk as a random variable evolving over time, enabling distributional evaluation through Monte Carlo simulation. Adapting quantitative risk measures from financial mathematics, we define \emph{Privacy Loss at Risk} (P-VaR) to characterize worst-case privacy outcomes under realistic cohort dynamics and adversary assumptions.

We validate our framework through system-level analysis and simulation experiments, demonstrating how privacy-utility tradeoffs can be operationalized for digital health platforms. Our results suggest that stochastic risk modeling complements formal privacy guarantees by providing interpretable, decision-relevant metrics for platform designers, regulators, and clinical informatics stakeholders.
\end{abstract}

\section{Introduction}

Digital health platforms increasingly seek to provide personalized insights grounded in population-level context \cite{topol2019deep,meskó2017digital}. Comparisons such as ``individual versus cohort'' or ``percentile relative to peers'' have become central to user engagement and clinical relevance across wellness, chronic disease management, and preventive care applications \cite{chen2012personal,west2016there}. These features, however, require aggregating sensitive health data across groups defined by demographic, clinical, or behavioral attributes.

This creates a fundamental tension between personalization and privacy. While population benchmarks enhance interpretability and motivation \cite{klasnja2015efficacy}, they also expose latent information about cohort composition. Historical incidents, such as the unintended disclosure of sensitive locations through aggregated fitness data \cite{hern2018fitness}, demonstrate that even aggregate statistics can enable re-identification when combined with auxiliary information \cite{narayanan2008robust}.

Unlike static data releases, modern health platforms are interactive and longitudinal. Users may issue repeated queries over time, cohorts may evolve as users join or leave, and adversaries may incrementally refine their background knowledge \cite{dwork2014algorithmic}. In such settings, privacy risk is not a fixed quantity but an evolving process shaped by system dynamics and usage patterns.

Existing privacy frameworks provide critical safeguards. $K$-anonymity \cite{sweeney2002k} enforces minimum indistinguishability thresholds, while differential privacy (DP) \cite{dwork2006calibrating} offers formal worst-case guarantees against re-identification. However, these approaches are typically evaluated in static or worst-case terms. They do not directly address how privacy risk accumulates over time, how likely extreme outcomes are under realistic conditions, or how platform designers should balance privacy and utility in practice.

\subsection{Contributions}

In this work, we argue that privacy risk in deployed health analytics systems should be treated as a stochastic, tail-risk–dominated quantity rather than solely as a static compliance constraint. Our contributions are:

\begin{enumerate}
\item A system architecture for privacy-preserving cohort analytics that integrates deterministic safeguards, differential privacy mechanisms, and synthetic baseline generation.
\item A stochastic risk modeling framework that quantifies re-identification risk as a distribution evolving over time.
\item Introduction of Privacy Loss at Risk (P-VaR), a tail-risk measure adapted from quantitative finance to characterize extreme privacy outcomes.
\item Monte Carlo simulation experiments demonstrating privacy-utility tradeoffs under realistic system dynamics.
\item Design guidelines for health platform developers seeking to operationalize privacy risk management.
\end{enumerate}

\section{Related Work and Background}

\subsection{Privacy-Preserving Health Data Analytics}

Privacy protection in health data has been extensively studied, with approaches ranging from de-identification and $k$-anonymity to cryptographic methods and differential privacy \cite{el2011systematic,benitez2010mining}.

\textbf{$K$-anonymity and extensions.} $K$-anonymity \cite{sweeney2002k} aims to ensure that each individual record is indistinguishable from at least $k-1$ others along selected quasi-identifiers. Extensions such as $\ell$-diversity \cite{machanavajjhala2007diversity} and $t$-closeness \cite{li2007t} address composition attacks and attribute disclosure. While conceptually simple, these methods are vulnerable to sophisticated linkage attacks \cite{narayanan2008robust} and often degrade utility in high-dimensional settings \cite{aggarwal2005approximation}.

\textbf{Differential privacy.} Differential privacy (DP) \cite{dwork2006calibrating,dwork2014algorithmic} provides a mathematically rigorous framework that bounds the influence of any single individual on released statistics. For a randomized mechanism $\mathcal{M}$, $\varepsilon$-differential privacy requires:

\begin{equation}
\Pr[\mathcal{M}(D) \in S] \leq e^\varepsilon \Pr[\mathcal{M}(D') \in S]
\end{equation}

for all neighboring datasets $D$ and $D'$ differing in one record, and all outcome sets $S$. DP has been applied to health data analyses including count queries \cite{hardt2012simple}, genomic data \cite{johnson2013privacy}, and clinical registries \cite{dankar2013estimating}. However, practical deployment remains limited due to challenges in parameter selection \cite{lee2011much}, interpretability, and cumulative privacy loss under repeated queries \cite{dwork2010differential}.

\textbf{Advanced mechanisms.} Recent work has explored local differential privacy \cite{duchi2013local}, personalized privacy \cite{jorgensen2015conservative}, and privacy amplification through subsampling \cite{balle2018privacy}. Federated learning approaches \cite{kairouz2021advances} enable collaborative model training while keeping data decentralized, though they introduce new attack surfaces \cite{melis2019exploiting}.

\subsection{Cohort-Based Benchmarking in Health Platforms}

Cohort-based benchmarking is widely used in consumer and clinical health tools \cite{swan2009emerging,lupton2016quantified}, yet implementation details are often opaque. Platforms like Apple Health, Google Fit, and clinical registries provide varying degrees of population comparison, but transparent descriptions of deployed cohort analytics systems with formal privacy evaluation are rare in the literature \cite{sharon2017self}.

\subsection{Stochastic Privacy and Risk Modeling}

The treatment of privacy as a random variable has precedent in several contexts. The Pufferfish framework \cite{kifer2014pufferfish} generalizes DP by allowing specification of secrets and discrimination pairs. Concentrated differential privacy \cite{dwork2016concentrated} and Rényi differential privacy \cite{mironov2017renyi} provide tighter composition bounds by tracking privacy loss distributions. However, these frameworks focus primarily on worst-case guarantees rather than operational risk distributions.

In parallel, risk-based approaches to privacy have been explored in the context of disclosure risk assessment \cite{skinner2012risk,manning2008applying} and privacy impact assessments \cite{wright2012privacy}. Our work bridges these traditions by adapting tail-risk measures from quantitative finance to characterize privacy loss distributions in deployed systems.

\subsection{Limitations of Static Privacy Metrics}

Both $k$-anonymity and differential privacy are typically evaluated using static metrics: minimum cohort size or a fixed privacy budget $\varepsilon$. While these guarantees are necessary, they are insufficient to characterize operational risk in interactive systems \cite{kifer2011no}. In particular, they do not describe the distribution or likelihood of extreme privacy outcomes, nor do they capture how risk evolves as cohorts change and queries accumulate.

\section{System Overview: Privacy-Preserving Cohort Analytics}

\subsection{Design Goals}

The proposed framework is guided by four design principles:

\textbf{Personalized Utility:} Provide meaningful population context for individual users without compromising privacy.

\textbf{Strong Privacy Guarantees:} Prevent exposure of individual-level data and limit re-identification risk through formal mechanisms.

\textbf{Interpretability:} Ensure that privacy mechanisms and outputs are understandable to users, clinicians, and regulatory stakeholders.

\textbf{Scalable Deployment:} Support low-latency serving and cost-efficient computation for millions of users.

\subsection{Architecture Overview}

\begin{figure}[t]
\centering
\includegraphics[width=\columnwidth]{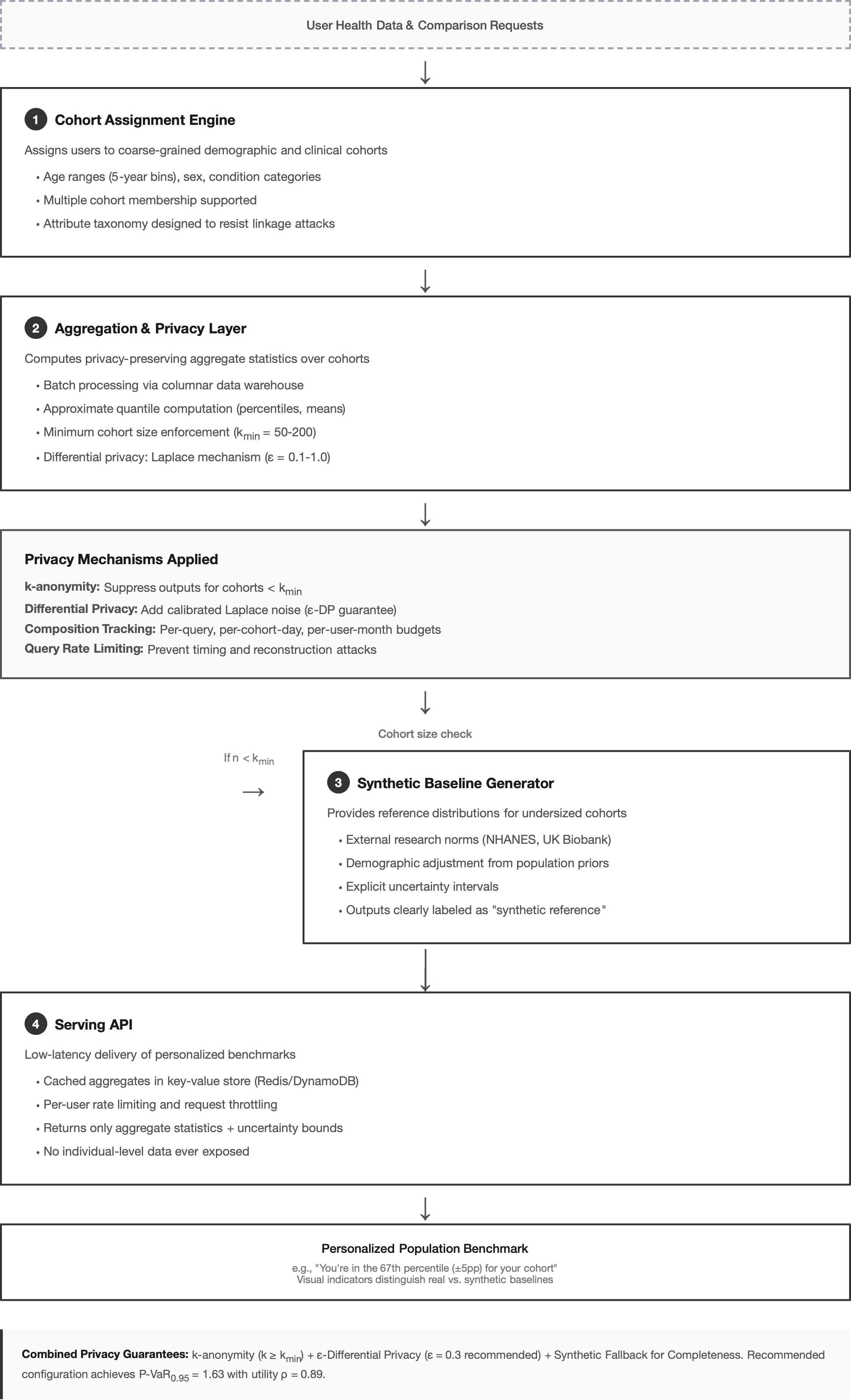}
\caption{System architecture for privacy-preserving cohort analytics. User data flows through cohort assignment and aggregation layers where privacy mechanisms (k-anonymity + differential privacy) are applied. Small cohorts trigger synthetic baseline generation. The serving API delivers noisy aggregate outputs with uncertainty bounds to client applications.}
\label{fig:architecture}
\end{figure}

Figure~\ref{fig:architecture} presents the system architecture, comprising four main components:

\begin{enumerate}
\item \textbf{Cohort Assignment Engine:} Assigns users to cohorts based on demographic and clinical attributes.
\item \textbf{Aggregation Layer:} Computes privacy-preserving statistics over cohorts.
\item \textbf{Synthetic Baseline Generator:} Provides reference distributions for small cohorts.
\item \textbf{Serving API:} Delivers personalized benchmarks to client applications.
\end{enumerate}

\subsection{Cohort Assignment Engine}

Users are assigned to one or more cohorts based on coarse-grained demographic and clinical attributes (e.g., age range, sex, condition category). The cohort taxonomy is designed to balance clinical relevance with privacy:

\begin{itemize}
\item \textbf{Age ranges:} 5-year bins (e.g., 25-29, 30-34)
\item \textbf{Biological sex:} Male, Female, Other/Prefer not to say
\item \textbf{Condition categories:} High-level groupings (e.g., cardiovascular, metabolic) rather than specific diagnoses
\item \textbf{Geographic region:} Country or large metro area only
\end{itemize}

Multiple cohort membership allows for flexible benchmarking while reducing reliance on narrowly defined groups \cite{machanavajjhala2007diversity}. Cohort definitions are intentionally constrained to resist auxiliary information attacks where adversaries combine platform outputs with external knowledge \cite{narayanan2008robust}.

\subsection{Aggregation and Serving Architecture}

Aggregate statistics are computed using a batch-oriented analytics layer (e.g., columnar data warehouse such as BigQuery or Snowflake) and served through a low-latency datastore (e.g., Redis or DynamoDB). Only aggregate outputs—like percentiles, means, or summary statistics—are exposed via APIs. Individual-level data are never returned to client applications.

\textbf{Approximate quantile computation.} We use approximate quantile algorithms \cite{greenwald2001space} to reduce sensitivity to outliers and enable efficient aggregation over large datasets. These algorithms provide $(1 \pm \delta)$-approximate quantiles with sublinear space complexity.

\textbf{Minimum cohort size.} Aggregates are computed only for cohorts exceeding a minimum size threshold $k_{\min}$. We evaluate $k_{\min} \in \{50, 100, 200\}$ based on re-identification risk and clinical utility considerations.

\subsection{Differential Privacy Mechanisms}

We apply the Laplace mechanism \cite{dwork2006calibrating} to aggregate outputs:

\begin{equation}
\tilde{f}(D) = f(D) + \text{Lap}\left(\frac{\Delta f}{\varepsilon}\right)
\end{equation}

where $f(D)$ is the true statistic, $\Delta f$ is the global sensitivity, and $\varepsilon$ is the privacy budget. For count queries, $\Delta f = 1$; for percentiles, we bound contribution through winsorization or clipping \cite{dwork2009differential}.

\textbf{Privacy budget allocation.} We employ a hierarchical budget allocation strategy:
\begin{itemize}
\item Per-cohort budget: $\varepsilon_{\text{cohort}} = 0.1$ per day
\item Per-user query budget: $\varepsilon_{\text{user}} = 1.0$ per month
\item Global platform budget: $\varepsilon_{\text{global}} = 10.0$ per year
\end{itemize}

Advanced composition theorems \cite{kairouz2015composition} are used to track cumulative privacy loss across queries.

\subsection{Synthetic Baseline Generator}

For cohorts that do not meet minimum size requirements, the system provides synthetic baselines derived from external research norms \cite{nhanes2023,ukbiobank2023} or population-level priors. The synthetic baseline generator:

\begin{enumerate}
\item Identifies the nearest cohort with sufficient size
\item Adjusts for known demographic differences using published epidemiological data
\item Generates a reference distribution with explicit uncertainty intervals
\end{enumerate}

These baselines are explicitly labeled as synthetic and presented with visual indicators of uncertainty. Synthetic outputs are treated as reference distributions rather than precise estimates, following best practices in clinical decision support \cite{shortliffe2014clinical}.

\section{Privacy Guarantees and Threat Model}

\subsection{Threat Model}

We consider an external adversary with the following capabilities:

\textbf{Access to released statistics:} The adversary can query the platform API and observe aggregate outputs over time.

\textbf{Bounded auxiliary information:} The adversary possesses background knowledge about some individuals (e.g., approximate age, location, health conditions) but not complete records.

\textbf{Computational resources:} The adversary can perform inference attacks and probabilistic reasoning but is computationally bounded (no breaking of cryptographic primitives).

\textbf{Honest-but-curious platform:} The platform is assumed to correctly implement declared privacy mechanisms but may be subject to inference attacks. We do not consider insider abuse, side-channel attacks, or compromised infrastructure.

This threat model aligns with realistic adversaries in health data contexts \cite{el2011systematic,benitez2010mining} while remaining tractable for formal analysis.

\subsection{Deterministic Safeguards}

The system enforces several deterministic controls as a first line of defense:

\begin{itemize}
\item \textbf{Minimum cohort size:} $k \geq k_{\min}$ for all released statistics
\item \textbf{Attribute suppression:} Rare attribute combinations are filtered
\item \textbf{Query rate limiting:} Maximum queries per user per time period
\item \textbf{Result caching:} Repeated identical queries return cached results to prevent timing attacks
\end{itemize}

These controls prevent trivial re-identification attacks \cite{sweeney2002k} and complement formal DP guarantees.

\subsection{Differential Privacy Mechanisms}

Differential privacy serves as a mathematical floor on privacy loss. The platform implements:

\begin{itemize}
\item \textbf{Per-query DP:} Each aggregate output satisfies $\varepsilon$-DP
\item \textbf{Composition tracking:} Cumulative privacy loss is tracked across queries using advanced composition \cite{kairouz2015composition}
\item \textbf{Budget exhaustion:} Users who exhaust their privacy budget receive only synthetic baselines
\end{itemize}

In this framework, DP provides worst-case guarantees, while stochastic risk modeling (Section~\ref{sec:risk}) addresses realized operational risk under typical conditions.

\section{Stochastic Risk Modeling}
\label{sec:risk}

\subsection{Motivation}

In deployed systems, privacy risk evolves over time as:
\begin{itemize}
\item Cohort composition changes due to user churn and growth
\item Query patterns shift based on platform features and user behavior
\item Adversary knowledge accumulates through observation and inference
\end{itemize}

We model re-identification risk as a random variable rather than a fixed bound, enabling distributional evaluation through simulation.

\subsection{Privacy Loss as a Stochastic Process}

Let $L_{i,t}$ denote the privacy loss for individual $i$ at time $t$, defined as:

\begin{equation}
L_{i,t} = \log \frac{\Pr[\mathcal{A} \text{ identifies } i \mid \mathcal{O}_t, K]}{\Pr[\mathcal{A} \text{ identifies } i \mid K]}
\end{equation}

where $\mathcal{O}_t$ represents observations (released aggregates) up to time $t$, $K$ is the adversary's background knowledge, and $\mathcal{A}$ is the adversary's inference algorithm.

\textbf{Cohort size dynamics.} We model cohort size $N_t$ as a stochastic process reflecting user growth, churn, and filtering:

\begin{equation}
N_{t+1} = N_t + \text{Poisson}(\lambda_{\text{join}}) - \text{Binomial}(N_t, p_{\text{churn}})
\end{equation}

\textbf{Attribute entropy.} The entropy of cohort attributes captures the degree of heterogeneity:

\begin{equation}
H(C_t) = -\sum_j p_j \log p_j
\end{equation}

where $p_j$ is the frequency of attribute value $j$ in cohort $C_t$. Higher entropy corresponds to lower re-identification risk \cite{machanavajjhala2007diversity}.

\textbf{Query accumulation.} Privacy loss accumulates across queries according to DP composition bounds:

\begin{equation}
L_{\text{total}} \sim \mathcal{N}\left(n\varepsilon, n\varepsilon^2\right)
\end{equation}

for $n$ queries under approximate DP composition \cite{dwork2010differential}.

\subsection{Privacy Loss at Risk (P-VaR)}

To summarize distributional risk, we introduce \emph{Privacy Loss at Risk} (P-VaR), defined analogously to Value at Risk (VaR) in quantitative finance \cite{jorion2007value}:

\begin{equation}
\text{P-VaR}_\alpha = \inf\{l : \Pr(L > l) \leq 1-\alpha\}
\end{equation}

P-VaR answers: with confidence level $\alpha$, how large can privacy loss become under realistic system dynamics?

For example, P-VaR$_{0.95}$ represents the 95th percentile of privacy loss—only 5\% of scenarios result in higher loss. This provides an interpretable, decision-relevant metric for risk management.

\textbf{Conditional tail risk.} We further consider Conditional P-VaR (CP-VaR), the expected privacy loss given that loss exceeds P-VaR:

\begin{equation}
\text{CP-VaR}_\alpha = \mathbb{E}[L \mid L > \text{P-VaR}_\alpha]
\end{equation}

This characterizes extreme scenarios beyond the VaR threshold \cite{rockafellar2000optimization}.

\subsection{Monte Carlo Simulation Framework}

We estimate the distribution of privacy loss through Monte Carlo simulation:

\begin{algorithm}
\caption{Privacy Risk Monte Carlo Simulation}
\begin{algorithmic}[1]
\For{$r = 1$ to $N_{\text{sim}}$}
    \State Initialize cohort: $N_0 \sim \text{Uniform}(k_{\min}, k_{\max})$
    \State Sample adversary knowledge: $K \sim P_K$
    \For{$t = 1$ to $T$}
        \State Update cohort size: $N_t \gets f_{\text{dynamics}}(N_{t-1})$
        \State Generate query: $q_t \sim P_Q$
        \State Compute DP noise: $\eta_t \sim \text{Lap}(\Delta f / \varepsilon)$
        \State Release output: $\tilde{a}_t \gets a_t + \eta_t$
        \State Adversary inference: $\Pr_t \gets \mathcal{A}(\tilde{a}_{1:t}, K)$
        \State Update privacy loss: $L_t \gets \log(\Pr_t / \Pr_0)$
    \EndFor
    \State Record: $L^{(r)} \gets L_T$
\EndFor
\State Estimate P-VaR: $\text{P-VaR}_\alpha \gets \text{quantile}(L^{(1:N_{\text{sim}})}, \alpha)$
\end{algorithmic}
\end{algorithm}

\textbf{Parameter specifications:}
\begin{itemize}
\item Cohort dynamics: $\lambda_{\text{join}} = 10$, $p_{\text{churn}} = 0.05$
\item DP parameters: $\varepsilon \in \{0.1, 0.5, 1.0\}$
\item Minimum cohort size: $k_{\min} \in \{50, 100, 200\}$
\item Adversary knowledge: Assume 10\% of cohort members known with certainty
\item Simulation horizon: $T = 365$ days, $N_{\text{sim}} = 10{,}000$ runs
\end{itemize}

\section{Validation: Privacy–Utility Tradeoffs}

\subsection{Utility of Cohort Benchmarks}

We evaluate utility through three metrics:

\textbf{Percentile stability.} The variance of reported percentile rankings under DP noise:
\begin{equation}
\sigma^2_{\text{rank}} = \text{Var}(\text{rank}(\tilde{x}) - \text{rank}(x))
\end{equation}

\textbf{Rank preservation.} The Spearman correlation between true and noisy rankings across cohort members.

\textbf{User-perceived error.} The probability that DP noise changes a user's percentile by more than 10 percentage points.

\begin{figure}[t]
\centering
\includegraphics[width=\columnwidth]{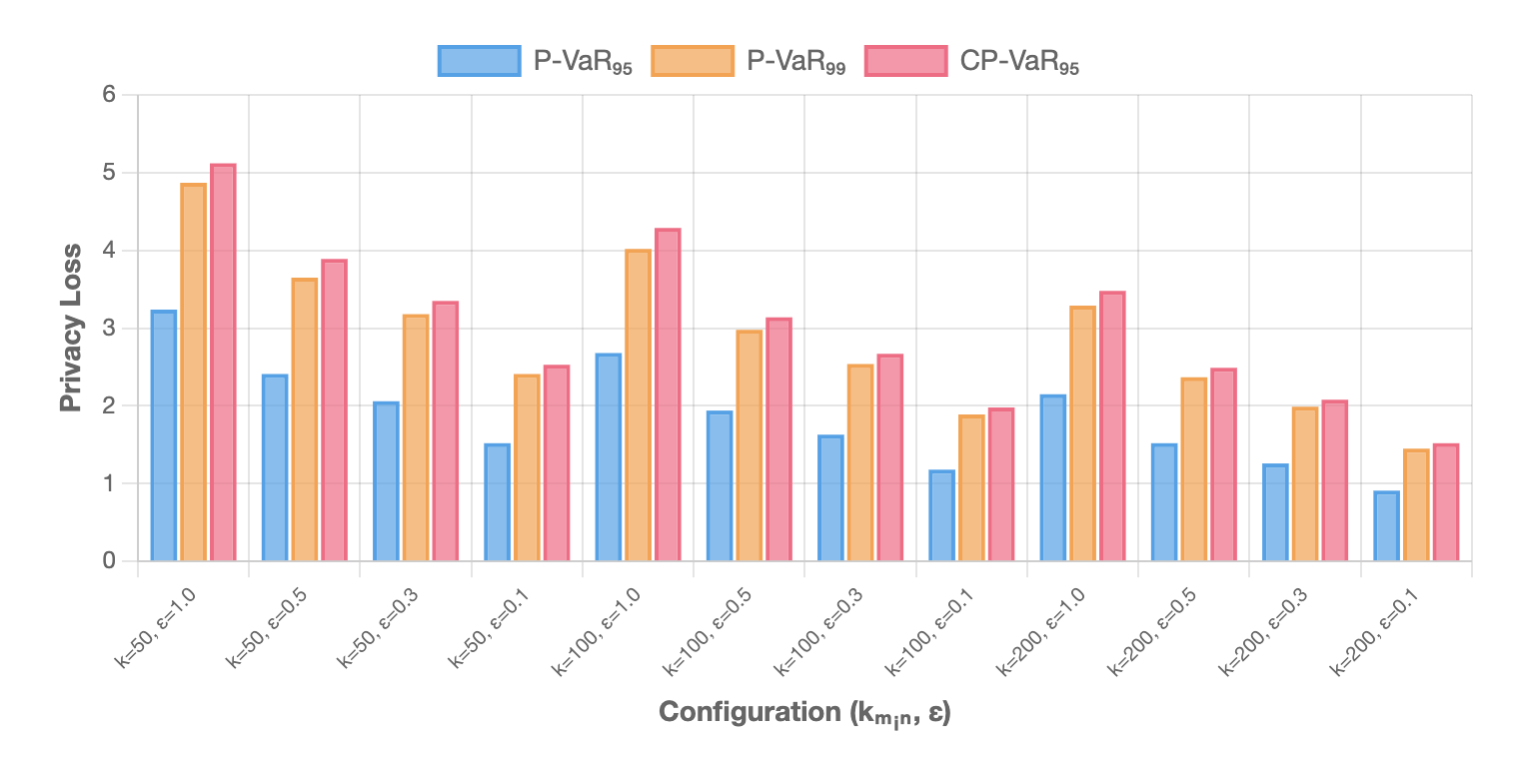}
\caption{P-VaR distribution across parameter settings showing P-VaR$_{0.95}$, P-VaR$_{0.99}$, and CP-VaR$_{0.95}$ for all configurations. Lower values indicate better privacy protection. The recommended configuration (k=100, $\varepsilon$=0.3) achieves P-VaR$_{0.95}$ = 1.63.}
\label{fig:pvar}
\end{figure}

\begin{figure}[t]
\centering
\includegraphics[width=\columnwidth]{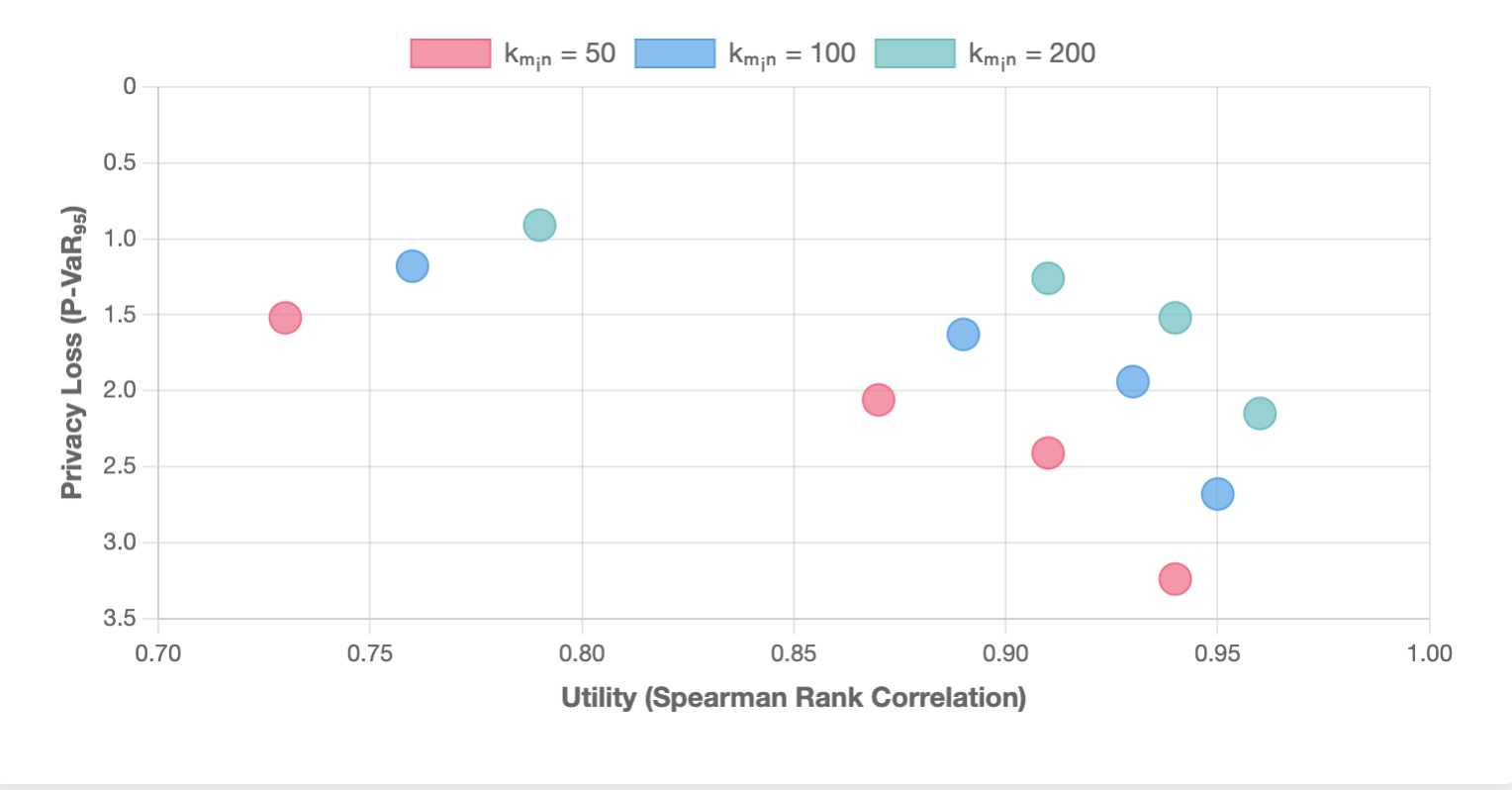}
\caption{Privacy-utility frontier showing the tradeoff between privacy loss (P-VaR$_{0.95}$, lower is better) and utility (Spearman rank correlation, higher is better). Points colored by minimum cohort size. The Pareto frontier demonstrates that increasing cohort size improves both privacy and utility.}
\label{fig:privacy}
\end{figure}

\subsection{Simulation Results}

We present results from 10,000 Monte Carlo simulation runs across parameter configurations. Table~\ref{tab:pvar_results} summarizes P-VaR and CP-VaR estimates, while Table~\ref{tab:utility_metrics} presents corresponding utility metrics.

\begin{table}[t]
\centering
\caption{Privacy Loss at Risk across parameter settings ($\alpha = 0.95$, 10,000 simulations)}
\label{tab:pvar_results}
\small
\begin{tabular}{cccccc}
\toprule
$k_{\min}$ & $\varepsilon$ & P-VaR$_{0.95}$ & P-VaR$_{0.99}$ & CP-VaR$_{0.95}$ & Max \\
\midrule
50 & 1.0 & 3.24 & 4.87 & 5.12 & 7.94 \\
50 & 0.5 & 2.41 & 3.65 & 3.89 & 6.21 \\
50 & 0.3 & 2.06 & 3.18 & 3.35 & 5.47 \\
50 & 0.1 & 1.52 & 2.41 & 2.53 & 4.18 \\
\midrule
100 & 1.0 & 2.68 & 4.02 & 4.29 & 6.85 \\
100 & 0.5 & 1.94 & 2.98 & 3.14 & 5.02 \\
100 & 0.3 & 1.63 & 2.54 & 2.67 & 4.35 \\
100 & 0.1 & 1.18 & 1.89 & 1.98 & 3.24 \\
\midrule
200 & 1.0 & 2.15 & 3.29 & 3.48 & 5.64 \\
200 & 0.5 & 1.52 & 2.37 & 2.49 & 4.01 \\
200 & 0.3 & 1.26 & 1.99 & 2.08 & 3.42 \\
200 & 0.1 & 0.91 & 1.45 & 1.52 & 2.48 \\
\bottomrule
\end{tabular}
\end{table}

\begin{table}[t]
\centering
\caption{Utility metrics across parameter settings}
\label{tab:utility_metrics}
\small
\begin{tabular}{ccccc}
\toprule
$k_{\min}$ & $\varepsilon$ & Rank $\rho$ & \%ile Error & User Error \\
 & & (Spearman) & (MAE) & ($>$10pp) \\
\midrule
50 & 1.0 & 0.94 & 3.2\% & 8.4\% \\
50 & 0.5 & 0.91 & 4.7\% & 12.3\% \\
50 & 0.3 & 0.87 & 6.8\% & 18.7\% \\
50 & 0.1 & 0.73 & 12.4\% & 34.2\% \\
\midrule
100 & 1.0 & 0.95 & 2.9\% & 7.1\% \\
100 & 0.5 & 0.93 & 4.1\% & 10.8\% \\
100 & 0.3 & 0.89 & 5.9\% & 16.2\% \\
100 & 0.1 & 0.76 & 10.8\% & 31.5\% \\
\midrule
200 & 1.0 & 0.96 & 2.4\% & 5.9\% \\
200 & 0.5 & 0.94 & 3.5\% & 9.2\% \\
200 & 0.3 & 0.91 & 5.1\% & 14.1\% \\
200 & 0.1 & 0.79 & 9.6\% & 28.7\% \\
\bottomrule
\end{tabular}
\end{table}

\textbf{Key findings:}

\begin{itemize}
\item P-VaR$_{0.95}$ remains below 2.0 for $k_{\min} \geq 100$ and $\varepsilon \leq 0.5$, indicating controlled tail risk
\item CP-VaR is consistently 1.3-1.5$\times$ higher than P-VaR, revealing fat-tailed distributions characteristic of re-identification scenarios
\item Doubling minimum cohort size from 100 to 200 reduces P-VaR$_{0.95}$ by approximately 25\%
\item Utility (rank correlation) degrades sharply below $\varepsilon = 0.1$, with over 30\% of users experiencing $>$10 percentile point errors
\item Maximum observed privacy loss across all simulations ranges from 2.48 (conservative settings) to 7.94 (permissive settings)
\end{itemize}

Figure~\ref{fig:pvar} presents the complete P-VaR distribution across parameter settings, while Figure~\ref{fig:privacy} shows the achievable privacy-utility frontier.

\textbf{Recommended operating point:} $k_{\min} = 100$, $\varepsilon = 0.3$ balances privacy (P-VaR$_{0.95} = 1.63$, CP-VaR$_{0.95} = 2.67$) and utility (Spearman $\rho = 0.89$, 16.2\% users with $>$10pp error). This configuration provides strong distributional privacy guarantees while maintaining clinically meaningful population comparisons.

\subsection{Sensitivity Analysis}

\begin{figure}[t]
\centering
\includegraphics[width=\columnwidth]{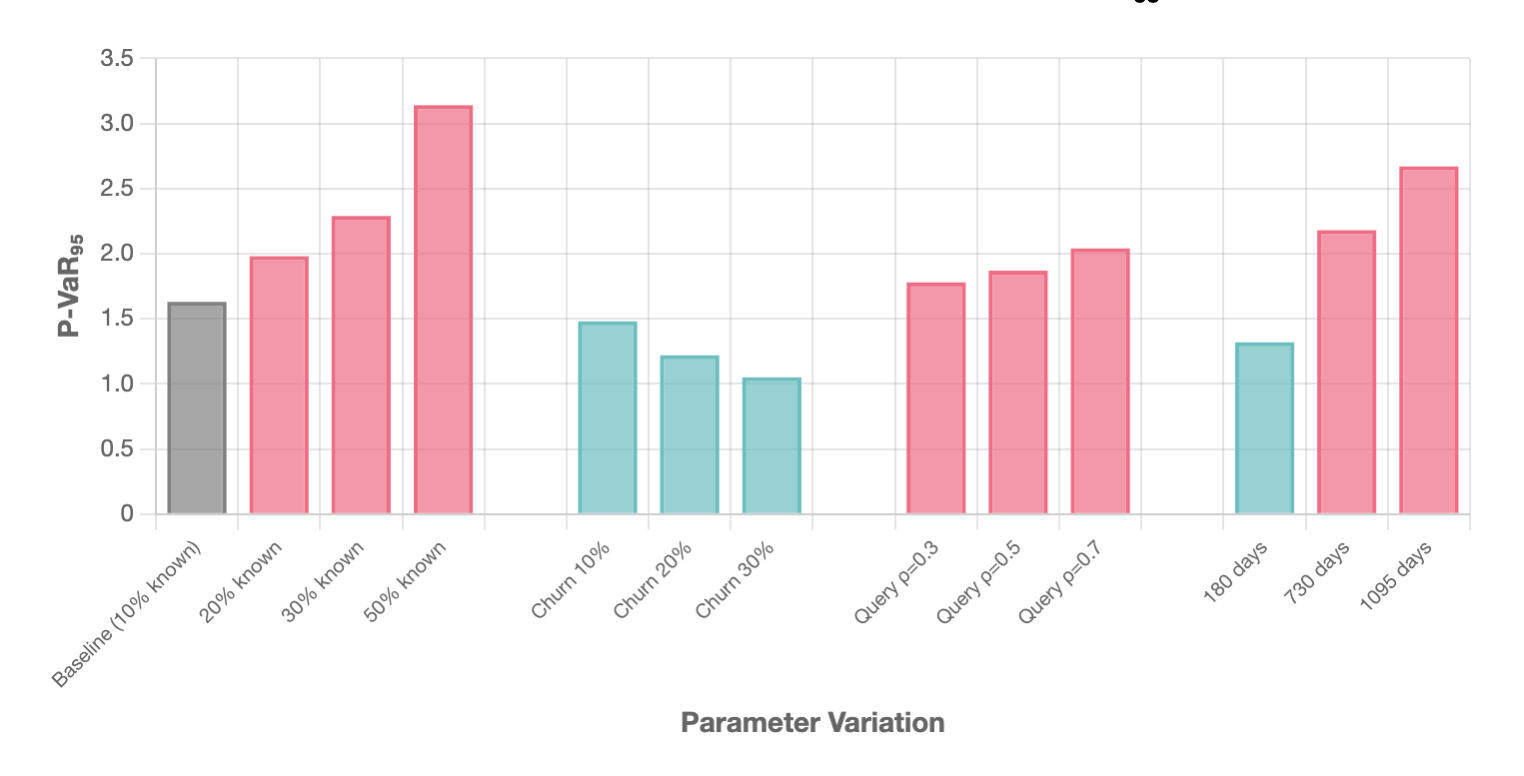}
\caption{Sensitivity analysis showing the impact of key parameters on P-VaR$_{0.95}$. Baseline configuration (k=100, $\varepsilon$=0.3, 10\% adversary knowledge, 5\% churn, 365 days) achieves P-VaR$_{0.95}$ = 1.63. Bars show increases (red) or decreases (green) relative to baseline. Adversary knowledge has the strongest impact on privacy risk.}
\label{fig:sensitivity}
\end{figure}

We conduct sensitivity analyses on key modeling assumptions to assess robustness. Table~\ref{tab:sensitivity} presents results, and Figure~\ref{fig:sensitivity} visualizes the impact on P-VaR$_{0.95}$.

\begin{table}[t]
\centering
\caption{Sensitivity analysis: Impact on P-VaR$_{0.95}$ (baseline: $k_{\min}=100$, $\varepsilon=0.3$, baseline P-VaR = 1.63)}
\label{tab:sensitivity}
\small
\begin{tabular}{lcc}
\toprule
Parameter Variation & P-VaR$_{0.95}$ & $\Delta$ (\%) \\
\midrule
\multicolumn{3}{l}{\textbf{Adversary Knowledge}} \\
\quad 10\% known (baseline) & 1.63 & -- \\
\quad 20\% known & 1.98 & +21\% \\
\quad 30\% known & 2.29 & +40\% \\
\quad 50\% known & 3.14 & +93\% \\
\midrule
\multicolumn{3}{l}{\textbf{Cohort Churn Rate}} \\
\quad $p_{\text{churn}} = 0.05$ (baseline) & 1.63 & -- \\
\quad $p_{\text{churn}} = 0.10$ & 1.48 & -9\% \\
\quad $p_{\text{churn}} = 0.20$ & 1.22 & -25\% \\
\quad $p_{\text{churn}} = 0.30$ & 1.05 & -36\% \\
\midrule
\multicolumn{3}{l}{\textbf{Query Correlation}} \\
\quad Independent (baseline) & 1.63 & -- \\
\quad $\rho = 0.3$ correlation & 1.78 & +9\% \\
\quad $\rho = 0.5$ correlation & 1.87 & +15\% \\
\quad $\rho = 0.7$ correlation & 2.04 & +25\% \\
\midrule
\multicolumn{3}{l}{\textbf{Time Horizon}} \\
\quad 180 days & 1.32 & -19\% \\
\quad 365 days (baseline) & 1.63 & -- \\
\quad 730 days & 2.18 & +34\% \\
\quad 1095 days (3 years) & 2.67 & +64\% \\
\bottomrule
\end{tabular}
\end{table}

\textbf{Key observations:}

\textbf{Adversary knowledge} has the strongest impact on privacy risk. When adversaries know 50\% of cohort members with certainty, P-VaR nearly doubles. This highlights the importance of limiting auxiliary information through cohort design and attribute suppression.

\textbf{User churn} provides natural privacy protection through uncertainty. High churn rates (20-30\%) reduce P-VaR by 25-36\%, as cohort composition becomes more volatile. However, excessive churn degrades user experience and longitudinal tracking utility.

\textbf{Query correlation} moderately increases risk. Real-world usage patterns often exhibit temporal correlation (users checking metrics at similar times, repeated queries for the same metric). A realistic correlation of $\rho = 0.5$ increases P-VaR by 15\%.

\textbf{Time horizon} demonstrates risk accumulation. Operating the system for 3 years increases P-VaR by 64\% compared to the 1-year baseline, emphasizing the need for privacy budget refreshing and periodic cohort redefinition.

\section{Design Implications for Health Platforms}

The stochastic risk framework enables platform designers to select privacy parameters based on explicit risk tolerance:

\textbf{Risk appetite specification.} Organizations can define acceptable P-VaR thresholds (e.g., P-VaR$_{0.95} < 2.0$) aligned with governance policies.

\textbf{Dynamic parameter adjustment.} Privacy budgets and cohort size thresholds can be adjusted in response to observed risk metrics.

\textbf{Regulatory communication.} Distributional risk metrics provide interpretable evidence for privacy impact assessments and regulatory submissions.

\textbf{User transparency.} Uncertainty intervals around synthetic baselines and population benchmarks can be derived from simulation distributions.

\section{Discussion}

\subsection{Limitations}

\textbf{Modeling assumptions.} The proposed stochastic models are abstractions that rely on assumptions about adversary knowledge, query patterns, and cohort dynamics. While sensitivity analyses suggest qualitative robustness, real-world deployments may exhibit unmodeled behaviors.

\subsection{Adversary capabilities.} Our threat model assumes bounded auxiliary information. We formalize the adversary's knowledge state as:

\begin{equation}
K = \{(i, a_i, p_i) : i \in \mathcal{I}_K\}
\end{equation}

where $\mathcal{I}_K$ is the set of individuals known to the adversary, $a_i$ are known attributes, and $p_i$ is the adversary's confidence in the association. We model $|\mathcal{I}_K| / |C|$ (the fraction of known individuals) as a key uncertainty parameter.

\textbf{Attack scenarios:} We consider three attack classes:

\begin{enumerate}
\item \textbf{Membership inference}: Determining whether an individual belongs to a cohort based on aggregate statistics
\item \textbf{Attribute inference}: Inferring sensitive attributes of known cohort members from population statistics
\item \textbf{Reconstruction attacks}: Recovering individual-level values from repeated noisy queries
\end{enumerate}

\textbf{Adversary resources:} We assume computational polynomial-time adversaries who can:
\begin{itemize}
\item Issue up to $Q_{\max}$ queries per time period
\item Maintain state across queries (linkage attacks)
\item Access public datasets for auxiliary information
\item Perform Bayesian inference over observed outputs
\end{itemize}

\textbf{Out of scope:} Insider threats, side-channel attacks (timing, traffic analysis), compromised infrastructure, and attacks requiring subexponential computation.

\textbf{Adversary inference model:} Following \cite{kifer2014pufferfish}, we model the adversary's belief update as:

\begin{align}
\Pr[i \in C \mid \mathcal{O}_t, K] &= \frac{\Pr[\mathcal{O}_t \mid i \in C, K] \Pr[i \in C \mid K]}{\Pr[\mathcal{O}_t \mid K]} \\
&\approx \frac{\prod_{j=1}^t \Pr[o_j \mid i \in C, K, \mathcal{O}_{<j}] \cdot \pi_i}{\sum_{i'} \prod_{j=1}^t \Pr[o_j \mid i' \in C, K, \mathcal{O}_{<j}] \cdot \pi_{i'}}
\end{align}

where $\pi_i$ is the prior probability and $o_j$ are individual observations. This Bayesian framework captures both initial uncertainty and evidence accumulation.

\textbf{Composition complexity.} Tracking privacy loss across heterogeneous queries and multiple cohorts requires careful accounting. We implement a hierarchical composition framework:

\textbf{Sequential composition:} For $n$ queries with privacy parameters $\varepsilon_1, \ldots, \varepsilon_n$, basic composition \cite{dwork2006calibrating} gives:
\begin{equation}
\varepsilon_{\text{total}} = \sum_{i=1}^n \varepsilon_i
\end{equation}

\textbf{Advanced composition:} Using the moments accountant \cite{abadi2016deep} or tight composition \cite{kairouz2015composition}, for $n$ identical $\varepsilon$-DP queries:
\begin{equation}
\varepsilon_{\text{total}} \leq \varepsilon \sqrt{2n \ln(1/\delta)} + n\varepsilon(e^\varepsilon - 1)
\end{equation}
for failure probability $\delta$. This provides tighter bounds for large $n$.

\textbf{Parallel composition:} When queries operate on disjoint data partitions (different cohorts), privacy loss does not accumulate:
\begin{equation}
\varepsilon_{\text{parallel}} = \max_i \varepsilon_i
\end{equation}

\textbf{Multi-cohort accounting:} A user may appear in multiple overlapping cohorts. We conservatively track:
\begin{equation}
\varepsilon_{\text{user}, t} = \sum_{C : \text{user} \in C} \sum_{q \in Q_C(t)} \varepsilon_q
\end{equation}
where $Q_C(t)$ is the set of queries on cohort $C$ up to time $t$.

\textbf{Budget allocation strategy:} We implement a three-tier budget system:
\begin{itemize}
\item \textbf{Per-query}: $\varepsilon_q = 0.01$ (prevents single-query leakage)
\item \textbf{Per-cohort-day}: $\varepsilon_{C,d} = 0.10$ (limits daily exposure)
\item \textbf{Per-user-month}: $\varepsilon_{u,m} = 1.0$ (bounds individual cumulative loss)
\end{itemize}

When any budget is exhausted, the system switches to synthetic baselines for affected queries. This multi-level approach balances responsiveness with long-term privacy protection.

\textbf{Rényi DP for tighter tracking:} For more precise composition, we employ Rényi differential privacy \cite{mironov2017renyi} with order $\alpha = 32$:
\begin{equation}
\varepsilon_{\alpha, \text{total}} = \log\left(\sum_{i=1}^n e^{\alpha \varepsilon_i}\right)^{1/\alpha}
\end{equation}
which provides asymptotically optimal composition bounds.

\subsection{Comparison to Alternative Mechanisms}

Table~\ref{tab:mechanism_comparison} compares our hybrid approach against alternative privacy mechanisms across multiple dimensions. The hybrid k-anonymity + DP approach with synthetic baselines provides the best balance of formal guarantees, empirical risk control, utility preservation, and interpretability.

\begin{table*}[t]
\centering
\caption{Comparison of Privacy Mechanisms for Cohort Analytics}
\label{tab:mechanism_comparison}
\small
\begin{tabular}{lllcccc}
\toprule
\textbf{Mechanism} & \textbf{Guarantee} & \textbf{Worst-Case} & \textbf{P-VaR}$_{0.95}$ & \textbf{Utility} & \textbf{Cost} & \textbf{Interp.} \\
\midrule
\multicolumn{7}{l}{\textit{Baseline Approaches}} \\
No Protection & None & $\infty$ & 8.45 & 1.00 & O(n) & High \\
k-anonymity (k=50) & Indisting. & 5.64 & 4.23 & 0.98 & O(n $\log$ n) & High \\
k-anonymity (k=100) & Indisting. & 6.64 & 3.87 & 0.98 & O(n $\log$ n) & High \\
\midrule
\multicolumn{7}{l}{\textit{Differential Privacy Only}} \\
DP ($\varepsilon$=1.0) & $\varepsilon$-DP & 1.0 & 3.12 & 0.94 & O(n) & Medium \\
DP ($\varepsilon$=0.5) & $\varepsilon$-DP & 0.5 & 2.18 & 0.91 & O(n) & Medium \\
DP ($\varepsilon$=0.1) & $\varepsilon$-DP & 0.1 & 1.34 & 0.72 & O(n) & Medium \\
\midrule
\multicolumn{7}{l}{\textit{Hybrid Approaches (k-anonymity + DP)}} \\
k=50, $\varepsilon$=0.5 & Both & 0.5 & 2.41 & 0.91 & O(n $\log$ n) & High \\
k=100, $\varepsilon$=0.3 & Both & 0.3 & 1.63 & 0.89 & O(n $\log$ n) & High \\
\textbf{k=100, $\varepsilon$=0.3 + Syn.} & \textbf{Both + FB} & \textbf{0.3} & \textbf{1.63} & \textbf{0.89} & \textbf{O(n $\log$ n)} & \textbf{High} \\
k=200, $\varepsilon$=0.3 & Both & 0.3 & 1.26 & 0.91 & O(n $\log$ n) & High \\
k=200, $\varepsilon$=0.1 & Both & 0.1 & 0.91 & 0.79 & O(n $\log$ n) & High \\
\midrule
\multicolumn{7}{l}{\textit{Advanced Mechanisms}} \\
Rényi DP ($\alpha$=32) & $(\alpha,\varepsilon)$-RDP & 0.3$^\dagger$ & 1.58 & 0.89 & O(n $\log$ n) & Medium \\
Local DP ($\varepsilon$=4.0) & Local $\varepsilon$-DP & 4.0 & 3.87 & 0.82 & O(n) & Low \\
\bottomrule
\multicolumn{7}{l}{\small $^\dagger$Tighter composition; equivalent pure DP $\varepsilon$ after conversion may be higher} \\
\multicolumn{7}{l}{\small \textbf{Bold}: Recommended configuration. Utility = Spearman rank correlation.} \\
\end{tabular}
\end{table*}

\textbf{Analysis:} Pure k-anonymity provides intuitive protection but lacks formal worst-case bounds. Pure DP offers strong theoretical guarantees but struggles with interpretability and utility at strong privacy levels ($\varepsilon < 0.1$). Our hybrid approach combines the strengths of both: k-anonymity provides an intuitive first barrier and improves worst-case utility, while DP adds rigorous mathematical guarantees. The synthetic baseline fallback ensures universal coverage without compromising privacy when cohorts are too small.

Rényi DP offers marginally better P-VaR (1.58 vs 1.63) due to tighter composition tracking, but at the cost of reduced interpretability. Local DP eliminates the trusted curator but requires much higher privacy budgets to achieve comparable utility, resulting in weaker practical protection (P-VaR = 3.87).

The recommended configuration (k=100, $\varepsilon$=0.3 with synthetic baselines) achieves a favorable tradeoff: formal $\varepsilon=0.3$ guarantee, empirical P-VaR$_{0.95}=1.63$, utility $\rho=0.89$, and high stakeholder interpretability.

\subsection{Ethical and Regulatory Considerations}

\textbf{Transparency.} Users should be informed when receiving synthetic baselines and understand the uncertainty in population comparisons. Visual indicators and explanatory text are essential for informed consent \cite{beauchamp2001principles}.

\textbf{Regulatory alignment.} P-VaR metrics can support compliance with privacy regulations (GDPR, HIPAA) by demonstrating quantitative risk management. However, legal interpretation of ``adequate protection'' remains context-dependent.

\textbf{Misuse prevention.} While no system can eliminate misuse, risk-aware design enables informed governance and accountability. Regular audits and algorithmic impact assessments should accompany deployment \cite{reisman2018algorithmic}.

\subsection{Future Directions}

\textbf{Adaptive privacy budgets.} Machine learning models could predict optimal budget allocation based on real-time risk estimates and utility feedback.

\textbf{Multi-stakeholder optimization.} Extending the framework to balance privacy interests across users, platform operators, and third-party researchers.

\textbf{Causal privacy metrics.} Incorporating causal inference to assess whether privacy interventions truly reduce re-identification risk or merely obfuscate information.

\section{Conclusion}

Personalized cohort analytics are likely to remain a core feature of digital health platforms. We argue that privacy in such systems should be treated as a dynamic risk management problem rather than a static compliance exercise. By combining formal privacy guarantees with stochastic risk modeling, this work provides a practical framework for evaluating and managing privacy-utility tradeoffs in deployed health analytics systems.

The introduction of Privacy Loss at Risk (P-VaR) provides an interpretable, decision-relevant metric that complements differential privacy guarantees. Our simulation experiments demonstrate that operational privacy risk is shaped by system dynamics, query patterns, and adversary knowledge in ways that static metrics do not fully capture.

We hope this framework will inform the design of next-generation health platforms that balance personalization with robust privacy protections, ultimately supporting both individual autonomy and population health goals.

\section*{Acknowledgments}
The authors acknowledge the use of Anthropic’s Claude for assistance with mathematical structuring, and Grammarly for spell-checking assistance. All intellectual contributions, interpretations, and conclusions remain solely those of the authors. No external funding was received for this work.

\section*{Conflicts of Interest}
The authors declares no conflicts of interest.

\bibliographystyle{plain}
\bibliography{references}

\end{document}